\documentclass[twocolumn,linenumbers]{aastex631}

\usepackage{threeparttable}

\begin{document}
\nolinenumbers
\title{ Insight-HXMT, NICER and NuSTAR views to the newly discovered black hole X-ray binary Swift J151857.0--572147}

\author[0000-0002-5554-1088]{Jing-Qiang Peng\textsuperscript{*}}
\email{pengjq@ihep.ac.cn}
\affiliation{Key Laboratory of Particle Astrophysics, Institute of High Energy Physics, Chinese Academy of Sciences, 100049, Beijing, China}
\affiliation{University of Chinese Academy of Sciences, Chinese Academy of Sciences, 100049, Beijing, China}
\author{Shu Zhang\textsuperscript{*}}
\email{szhang@ihep.ac.cn}
\affiliation{Key Laboratory of Particle Astrophysics, Institute of High Energy Physics, Chinese Academy of Sciences, 100049, Beijing, China}

\author[0000-0001-5160-3344]{Qing-Cang Shui\textsuperscript{*}}
\email{shuiqc@ihep.ac.cn}
\affiliation{Key Laboratory of Particle Astrophysics, Institute of High Energy Physics, Chinese Academy of Sciences, 100049, Beijing, China}
\affiliation{University of Chinese Academy of Sciences, Chinese Academy of Sciences, 100049, Beijing, China}

\author[0000-0001-8768-3294]{Yu-Peng Chen}
\affiliation{Key Laboratory of Particle Astrophysics, Institute of High Energy Physics, Chinese Academy of Sciences, 100049, Beijing, China}

\author[0000-0001-5586-1017]{Shuang-Nan Zhang}
\affiliation{Key Laboratory of Particle Astrophysics, Institute of High Energy Physics, Chinese Academy of Sciences, 100049, Beijing, China}
\affiliation{University of Chinese Academy of Sciences, Chinese Academy of Sciences, 100049, Beijing, China}

\author[0000-0003-3188-9079]{Ling-Da Kong}
\affiliation{Institute f{\"u}r Astronomie und Astrophysik, Kepler Center for Astro and Particle Physics, Eberhard Karls, Universit{\"a}t, Sand 1, D-72076 T{\"u}bingen, Germany}
\author{A. Santangelo}
\affiliation{Institute f{\"u}r Astronomie und Astrophysik, Kepler Center for Astro and Particle Physics, Eberhard Karls, Universit{\"a}t, Sand 1, D-72076 T{\"u}bingen, Germany}
\author{Zhuo-Li Yu}
\affiliation{Key Laboratory of Particle Astrophysics, Institute of High Energy Physics, Chinese Academy of Sciences, 100049, Beijing, China}
\author[0000-0001-9599-7285]{Long Ji}
\affiliation{School of Physics and Astronomy, Sun Yat-Sen University, Zhuhai, 519082, China}

\author[0000-0002-6454-9540]{Peng-Ju Wang}
\affiliation{Institute f{\"u}r Astronomie und Astrophysik, Kepler Center for Astro and Particle Physics, Eberhard Karls, Universit{\"a}t, Sand 1, D-72076 T{\"u}bingen, Germany}

\author[0000-0003-4856-2275]{Zhi Chang}
\affiliation{Key Laboratory of Particle Astrophysics, Institute of High Energy Physics, Chinese Academy of Sciences, 100049, Beijing, China}
\author{Jian Li}
\affiliation{CAS Key Laboratory for Research in Galaxies and Cosmology, Department of Astronomy, University of Science and Technology of China, Hefei 230026, China}
\affiliation{School of Astronomy and Space Science, University of Science and Technology of China, Hefei 230026, China}
\author[0000-0003-2310-8105]{Zhao-sheng Li}
\affiliation{ Key Laboratory of Stars and Interstellar Medium, Xiangtan University, Xiangtan 411105, Hunan, China}



\begin{abstract}
\nolinenumbers
The systematic properties are largely unknown for the black hole X-ray binary Swift J151857.0--572147 newly discovered in the 2024 outburst. The nature of a black hole can be completely defined by specifying the mass and dimensionless spin parameter. Therefore, accurate measurement of the two fundamental parameters is important for understanding the nature of black holes.
The joint spectral fitting of a reflection component with simultaneous observations from Insight-HXMT, NICER and NuSTAR reveals for the first time a black hole  dimensionless spin of $0.84^{+0.17}_{-0.26}$ and an inclination angle of $21.1^{+4.5}_{-3.6}$ degree for this system. Monitoring of the soft state by NICER results in disk flux and temperature following $F_{\rm disk} \propto T_{\rm in}^{3.83\pm 0.17}$.  For the standard thin disk,
$L_{\rm disk}\approx 4\pi R_{\rm in}^{2}\sigma T_{\rm in}^{4}$, so the relationship between the flux and temperature of the disk we measured indicates that the inner radius of the disk is stable and the disk is in the Innermost Stable Circular Orbit. With an empirical relation built previously between the black hole outburst profile and the intrinsic power output, the source distance is estimated 
as $5.8\pm 2.5$ kpc according to the outburst profile and peak flux observed by Insight-HXMT and NICER. Finally, a black hole mass of $3.67\pm1.79-8.07\pm 4.20 M_\odot$ can be inferred from a joint diagnostic of the aforementioned parameters measured for this system. This system is also consistent with most black hole X-ray binaries with high spin and a mass in the range of 5--20 $M_\odot$

\end{abstract}



\section{Introduction} \label{sec:intro}
Black hole (BH) low-mass X-ray binaries consist of a companion star with a mass less than $1 M_\odot$ and a BH.
Most of the BH low-mass X-ray binaries can only be observed by X-ray telescopes when they enter into outbursts,  during which their different spectral states can be observed  and traced in the Hardness-Intensity diagram (HID)  \citep{2001Homan, 2004Fender, 2012Motta}. 
For the Low Hard State (LHS), the emission is dominated by non-thermal emission from the corona/jet, accompanied by a relatively low percentage of thermal emission from the disk \citep{2005Belloni}.
The accretion disk is commonly thought to be truncated, but as of now, there is no definitive observational evidence that the accretion disk is truncated in the LHS \citep{2006Miller,2013Reynolds,2024Draghis}. As the accretion rate increases, the source passes through the intermediate states (hard and soft intermediate states) to the High Soft State (HSS).
In the HSS, where the inner edge of the disk is expected to lie at the  Innermost Stable Circular Orbit (ISCO) and dominate the emission in the form of thermal. The non-thermal emission in HSS is relatively weak, usually accounting for less than 25\% of the total \citep{1997Esin,2008Gierlinski}.

The nature of a black hole can be completely defined by specifying two parameters:  mass $M$ and dimensionless spin $a$. Therefore, accurate measurement of the two fundamental parameters is important for understanding the nature of black holes.
The spin of a black hole fundamentally changes the geometry of spacetime.  The gravitational well of an extreme Kerr black hole of the same mass is significantly deeper compared to a spinless black hole, resulting in a much harder X-ray spectrum and significantly enhancing its efficiency in converting the accreted rest mass into radiant energy. And the spin of black holes plays an important role in the mechanism of jets, core collapse of gamma-ray bursts, gravitational-wave astronomy in predicting the waveforms of merging black holes, and so on. \citep{1977MNRAS.179..433B,1993ApJ...405..273W,1997Zhang,2006ApJ...652..518M,2006PhRvD..74d1501C,2011MNRAS.416..941S}
The mass of a black hole provides a physical scale. The mass distribution of such black holes can provide important clues to the end stages of the evolution of massive stars.

The BH spin can be inferred from fitting the continuum and reflection spectra. The continuum spectrum approach requires an inner disk to stay around ISCO, independent mass, inclination, and distance measurements, and an assumption about the color correction factor,
and many models can be used to estimate BH spin e.g. {\tt kerrbb, kerrbb2} \citep{1997Zhang,2005Li},
while the reflection spectrum,  approach deals with the broadened iron line and the Compton hump around 20 keV \citep{2006Brenneman,2009Miller},
commonly used models are the {\tt relxill} family, etc.\citep{2016Dauser}.
The BH mass measurement is usually a tough task. The most reliable BH mass estimation comes from the mass function obtained from observing the orbital modulation of the line emission from the companion \citep{1996ApJ...470L..61K,2016ApJS..222...15T}. BH mass can also be constrained from the merger of the compact objects, and BHs discovered by the Laser Interferometer Gravitational-Wave Observatory and its Virgo interferometer (LIGO/Virgo) cover a rather wide mass range of $3-100 M_\odot$ \citep{2022Msngr.186....3B}. 
For other approaches, \cite{2023P,2024ApJ...965L..22P}  reported the BH mass of SLX 1746--331 by assuming its HSS reaches 30\% $L_{\rm Edd}$, which results in a BH mass similar to that derived from an empirical relation of BH mass with disk temperature.

Swift J151857.0--572147 is a new black hole X-ray binary first observed by SWIFT/BAT.  A spectral fit to the combined Swift/XRT data shows a power-law slope with a photon index of 1.78$\pm$0.02 \citep{2024ATel16500....1K}.
\cite{2024ATel16503....1C} found that the radio spectrum of the source is inverted with a preliminary spectral index of $\alpha \sim$ +0.5, where flux density as a function of frequency ($f_{\nu}$) is proportional to $\nu^{\alpha}$. This inverted index and the reported X-ray photon index of +1.78 are consistent with an X-ray binary in the hard state, suggesting that Swift J151857.0--572147 is either a black hole or (radio-bright) neutron star X-ray binary.
An extremely bright radio flare was detected by ATCA On March 9,  which is usually suggestive of having discrete ejecta from a BH X-ray binary during the transition between hard and soft X-ray states \citep{2024Car}.
Subsequently, Swift/XRT observation of Swift J151857.0--572147 confirmed a likely transition to the soft state on March 10 \citep{2024Del}.
The HI spectroscopic distance constraints for Swift J151857.0--572147 give a rather large range of 4.48--15.64 kpc \citep{2024B}. 
By taking a distance of 10 kpc, \cite{2024M} fitted the joint IXPE and NuSTAR spectra and reported rather broad parameter coverages: $0.6 \pm 0.2-0.7\pm0.2$ for BH spin, $38 \pm 9$°$-$$47 \pm 15$° for disk inclination, $9.2 \pm 1.6 - 10.5 \pm 1.8 M_\odot$ for BH mass.
 \cite{2024arXiv240617629C} detected type C quasi-periodic oscillations (QPOs) in Swift J151857.0--572147 with Insight-HXMT data.

With the observations from Insight-HXMT, NICER, and NuSTAR we perform the first detailed investigation upon the systematic properties of Swift J151857.0--572147.
In Section \ref{obser}, we describe the observations and data reduction. The detailed results are presented in Section \ref{result}. The results are then discussed, and the conclusions are presented in Section \ref{diss}.

\section{Observations and Data reduction}
\label{obser}

\begin{table}[]
    \centering
		\caption{NICER observations of Swift J151857.0--572147 during the 2024 outburst. }
		\begin{tabular}{cccc}
		%\begin{tabular}{|c|m{2cm}|}
		  \hline
              \hline
        NICER & Observed date & Exposure Time 
         \\ ObsID &(MJD)&(s)
       \\ \hline
        7204220111 &60387.55& 4193\\ 
        7661010101 &60404.18 & 3907\\ 
        7204220112 &60405.22 & 569 \\ 
       7204220113 &60408.90 & 228  \\ 
       7204220114 &60409.16 & 404  \\ 
       7661010103 &60412.06 & 2682 \\ 
       7204220115 &60412.77 & 262 \\ 
       7204220116 &60413.67 & 799 \\ 
      7204220117 &60414.06 & 1411  \\ 
      7204220118 &60415.41 & 1065 \\ 
      7204220119 &60416.25 & 1265\\ 
       7204220120 &60417.29 & 565 \\ 
       7661010104 &60417.41 & 3218 \\ 7661010105 &60425.09 & 2474 \\ 
       7661010107 &60433.02 & 2021 \\

    \hline
        \label{nicerobservation}    

    \end{tabular}

\end{table}

\subsection{NICER}

NICER  was launched by the Space X Falcon 9 rocket on 3 June 2017 \citep{2016Gendreau}. NICER has a large effective area and high temporal resolution in soft X-ray band (0.2--12 keV),  which may allow us to study black body components at low temperatures better. We selected NICER observations during the soft state of Swift J151857.0--572147, covering a time zone from MJD 60387--MJD 60433.

We use the official software HEASOFT V6.33/NICERDAS v012 with the latest CALDB xti20240216 for NICER data analysis.
NICER data are reduced using the standard pipeline tool {\tt nicer}l2\footnote{\url{https://heasarc.gsfc.nasa.gov/lheasoft/ftools/headas/nicerl2.html}}. 
We extract light curves using {\tt nicer}l3-lc\footnote{\url{https://heasarc.gsfc.nasa.gov/docs/software/lheasoft/ftools/headas/nicerl3-lc.html}} in 1--4 keV, 4--10 keV and 1--10 keV.
To extract the spectrum, we utilize {\tt nicer}l3-spect\footnote{\url {https://heasarc.gsfc.nasa.gov/docs/software/lheasoft/help/nicerl3-spect.html}}, employing the "{\tt nibackgen3C50}\footnote{\url{https://heasarc.gsfc.nasa.gov/docs/nicer/analysis_threads/background/}}" model to estimate the background for spectral analysis.
We select an energy range of 2--10 keV for spectrum fitting.
Additionally, {\tt nicer}l3-spect automatically applies the systematic error using {\tt niphasyserr}. In the energy range of 0.3--10 keV, the systematic error is about 1.5\%.

\subsection{NuSTAR}

NuSTAR is the first mission to employ two focusing techniques for detecting hard X-rays at 3--79 keV. It was launched on June 13, 2012, at 9 am PDT. \citep{2013Harrison}. 
As shown in Table \ref{observation}, NuSTAR conducted two observations of Swift J151857.0--572147.
For NuSTAR data analysis, we employ the software HEASOFT V6.33/NuSTARDAS v2.1.2 with the calibration database (CALDB 20230613) to reduce the data.
We extract NuSTAR-filtered data using the standard pipeline program {\tt nupipeline},  
The spectrum is extracted from a 120$''$ circle region centered on the source and the background is generated from a 60$''$ circle region away from the source. We utilized FPMA and FPMB 3--78 keV data for spectral analysis.

\subsection{Insight-HXMT}
Insight-HXMT, the first Chinese X-ray astronomy satellite, was successfully launched on June 15, 2017 \citep{2014Zhang, 2018Zhang, 2020Zhang}. The satellite is equipped with three scientific payloads: the low-energy X-ray telescope (LE) featuring an SCD detector operating in the 1--15 keV range with an effective area of 384 $\rm cm^{2}$ \citep{2020Chen}, the medium-energy X-ray telescope (ME) equipped with a Si-PIN detector covering 5--35 keV with an effective area of 952 $\rm cm^{2}$ \citep{2020Cao}, and the high-energy X-ray telescope (HE) utilizing a phoswich NaI(CsI) detector for the 20--250 keV energy range with an effective area of 5100 $\rm cm^{2}$ \citep{2020Liu}.

The Insight-HXMT began observing Swift J151857.0--572147 on March 4, 2024 and ended on March 17. However, due to contamination by Cir X--1 which is located in the field of view at a position close to Swift J151857.0--572147,  only the Insight-HXMT HE and ME observations simultaneous with NICER are adopted for joint spectral analysis (Table \ref{observation}). This is because during the joint observations, Cir X--1 was in dip phase with a rather soft energy spectrum and hence the expected count rates at ME and HE bands are rather small. 

We extract the data from LE ME and HE using the Insight-HXMT Data Analysis software {\tt{HXMTDAS v2.06}}. 
The data are filtered with the criteria recommended by the Insight-HXMT Data Reduction Guide {\tt v2.06} \footnote{\url{http://hxmtweb.ihep.ac.cn/SoftDoc/648.jhtml}}.
Due to the relatively high impact of Cir X--1 on LE, the energy bands considered for spectral analysis are ME 8--28 keV and HE 28--100 keV. One percent systematic error is added to data \citep{2020Liao}, and errors are estimated via  Markov Chain Monte-Carlo (MCMC) chains with a length of 20000.

\section{Results}
\label{result}

\subsection{Light curve and Hardness-intensity diagram}
\label{light curve}

\begin{figure*}
	\centering
	\includegraphics[angle=0,scale=0.6]{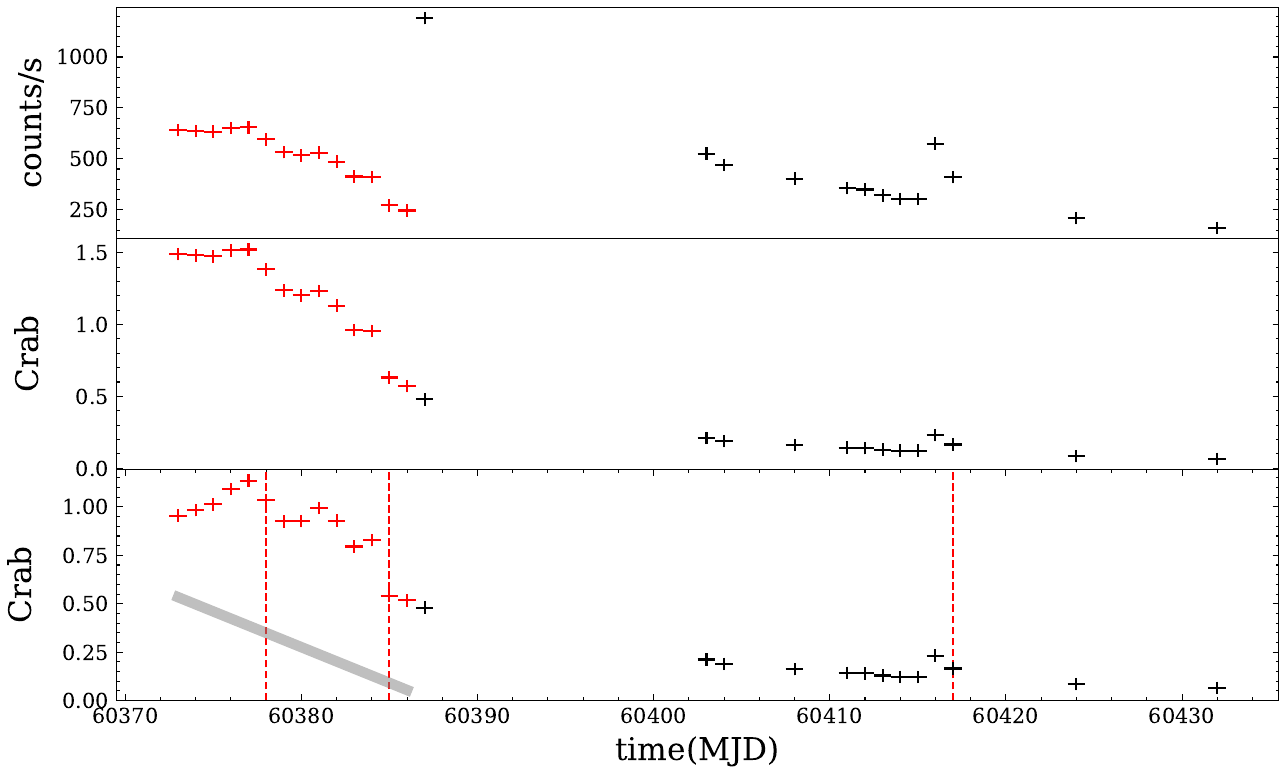}
	\caption{The light curves of Swift J151857.0--572147 observed by Insight-HXMT and NICER during the 2024 outburst. The red dots represent Insight-HXMT observations and the black dots represent NICER observations. Top panel: the light curve of Insight-HXMT LE and NICER in 2--12 keV. Middle panel: The light curves are normalized to the Crab. Bottom panel: The light curve of Insight-  HXMT after subtracting off the  Cir X--1 (a grey area) and the light curve of NICER. The red dotted lines correspond to about 90\%, 50\%, and 10\% of the peak flux, respectively.}
	\label{lcurve}
\end{figure*}

\begin{figure}
	\centering
	\includegraphics[angle=0,scale=0.5]{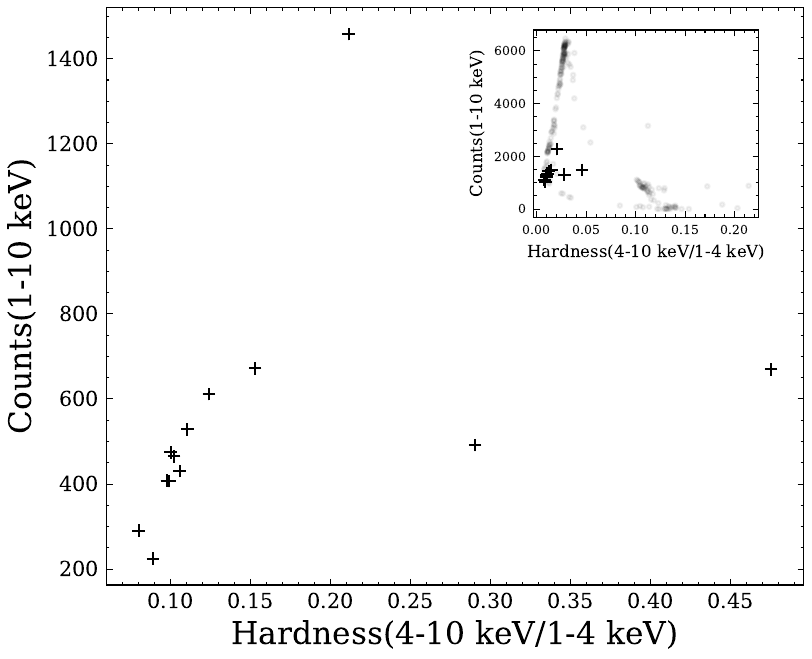}
	\caption{The NICER HID of Swift J151857.0--572147, where the hardness is defined as the ratio of 4--10 keV to 1--4 keV count rate.  The subplot shows the distribution of the outburst trajectory of Swift J151857.0--572147 after scaling the count rate relative to the trajectory of the 2021 outburst of GX 339--4.}
	\label{HID}
\end{figure}

Figure \ref{lcurve} shows the light curves of Swift J151857.0–-572147 observed by Insight-HXMT and NICER during the 2024 outburst.
The red dots represent Insight-HXMT observations and the black dots represent NICER observations.
We normalized the light curves to the Crab Nebula by using the count rate ratios between Insight-HXMT and NICER.
Since Swift J151857.0--572147 is 0.3 degrees away from Cir X--1, both sources are enclosed in the field of view of Insight-HXMT. Fortunately, Cir X--1 has a relatively stable profile for flux evolution over its orbital period of 16.6 days, the overall flux evolution trend at the LE band can be constructed with Insight-HXMT observations outside the outburst of 
Swift  J151857.0--572147.
So we extract the Cir X--1 LE light curve at LE 2-12 keV, then normalize it with the counts of Crab in the same energy band and fit it to get the flux evolution trend. (see the grey shading in the bottom panel of Figure \ref{lcurve}). Then we subtract off the Cir X--1 and get the LE light curve of the Swift J151857.0--572147 at 2--12 keV.

The HID is built with NICER light curves in  1--4, 4--10, and 1--10 keV.  The hardness is defined as the count rate ratio of 4--10 keV to 1--4 keV, while the intensity takes the count rate of 1--10 keV.  As shown in Figure \ref{HID}, the source was in a soft state during almost the whole NICER observations.

\subsection{The spectral analysis}
\label{parameters}

\begin{table}[htbp]
    \setlength{\tabcolsep}{1pt}
    \centering
		\caption{The quasi-simultaneous NICER, NuSTAR and Insight-HXMT observations of Swift J151857.0--572147  during the 2024 outburst. }
		\begin{tabular}{cccccc}
            
		%\begin{tabular}{|c|m{2cm}|
		  \hline
		   \hline
		   NICER & Observed date & Exposure Time \\ObsID& (MJD) & (s) \\ \hline
      7204220111 &60387.5 & 4193 \\ 
       \hline \hline
                Insight-HXMT &  Observed date & Exposure time &Exposure time\\
         &  & ME& HE 
         \\ ObsID& (MJD) & (s) & (s) 
       \\ \hline
       P061437400802 & 60386.7& 3704 & 1925  \\ \hline \hline
         NuSTAR &  Observed date & Exposure time & Exposure time\\
         &  & FPMA & FPMB 
         \\ ObsID& (MJD) & (s) & (s) 
         \\ \hline
       91001311002 &  60386.7 &13880& 14300 \\ 
       91001311004 &  60387.6 &9195& 9449 \\  
    \hline
        \label{observation} &     

    \end{tabular}

\end{table}

\begin{figure}
	\centering
	\includegraphics[angle=0,scale=0.3]{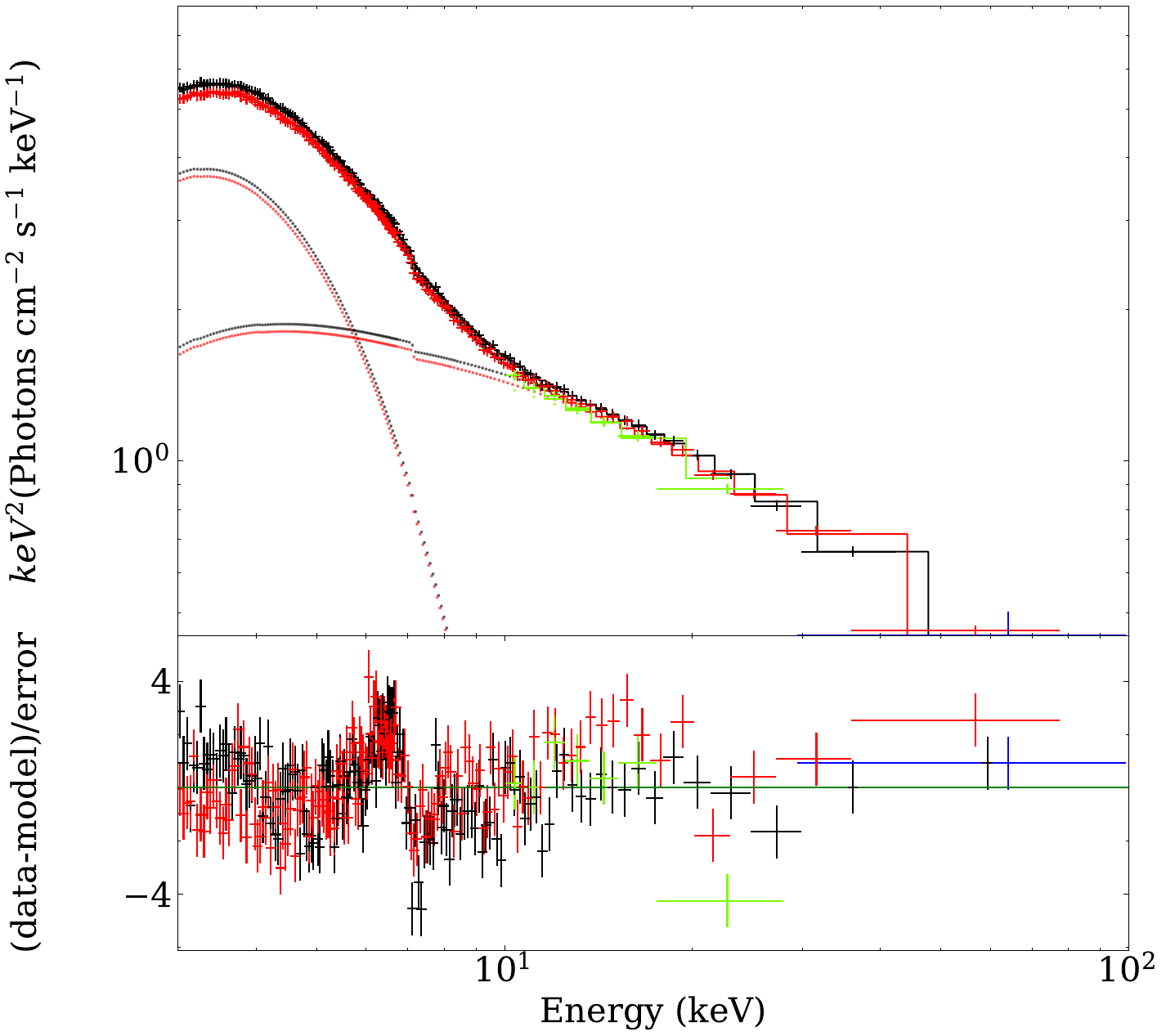}
	\caption{The simultaneous broadband spectrum with the model ({\tt constant*tbabs*(diskbb+cutoffpl)}) of Swift J151857.0--572147 is observed from  NuSTAR/FPMA (black), NuSTAR/FPMB (red), Insight-HXMT ME (green) and Insight-HXMT HE (blue).}
	\label{re}
\end{figure}

\begin{figure}
	\centering
	\includegraphics[angle=0,scale=0.3]{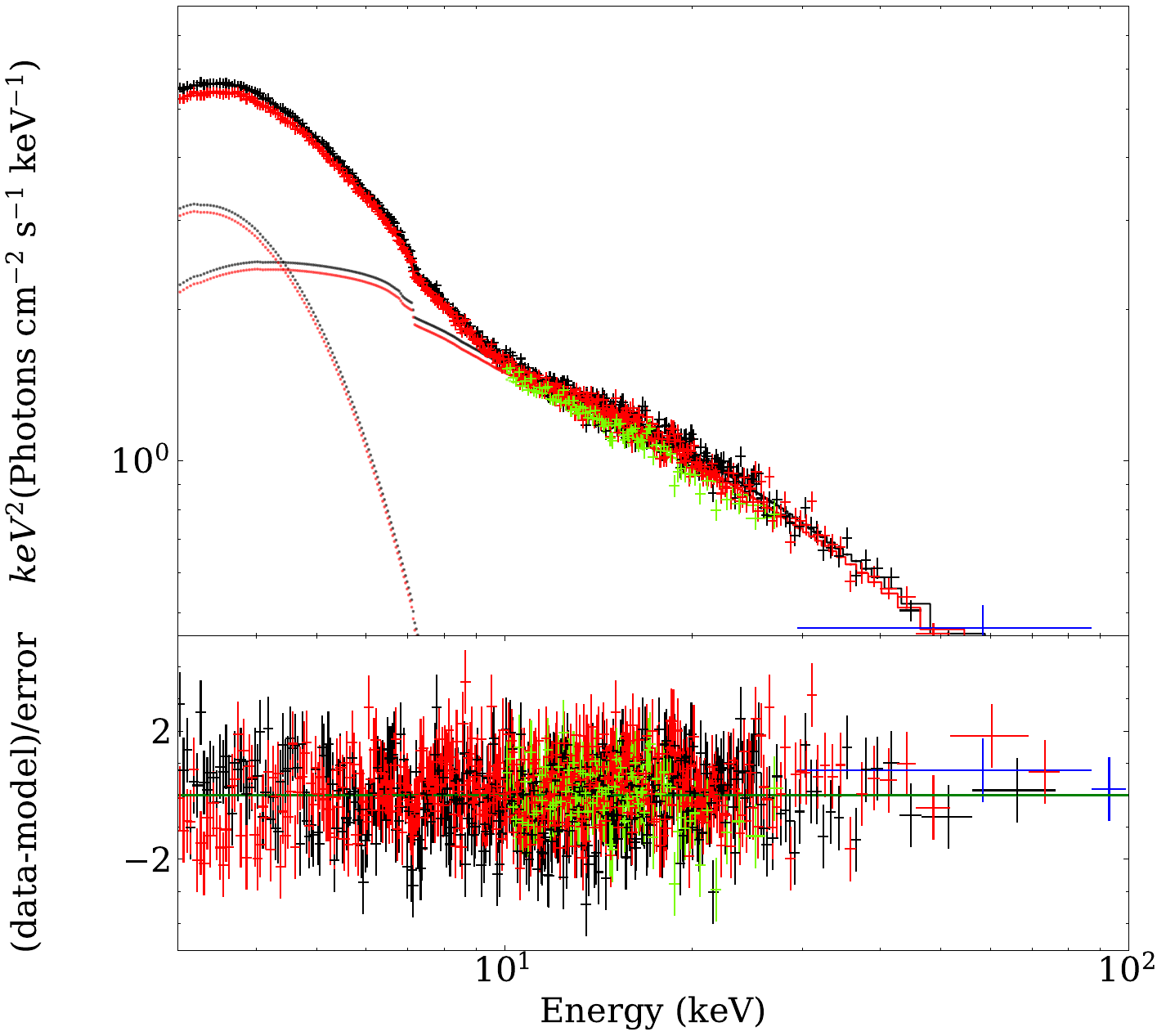}
	\caption{The simultaneous broadband spectrum with the model M1 of Swift J151857.0--572147 is observed from  NuSTAR/FPMA (black), NuSTAR/FPMB (red), Insight-HXMT ME (green) and Insight-HXMT HE (blue).}
	\label{nuhx}
\end{figure}

\begin{figure}
	\centering
	\includegraphics[angle=0,scale=0.3]{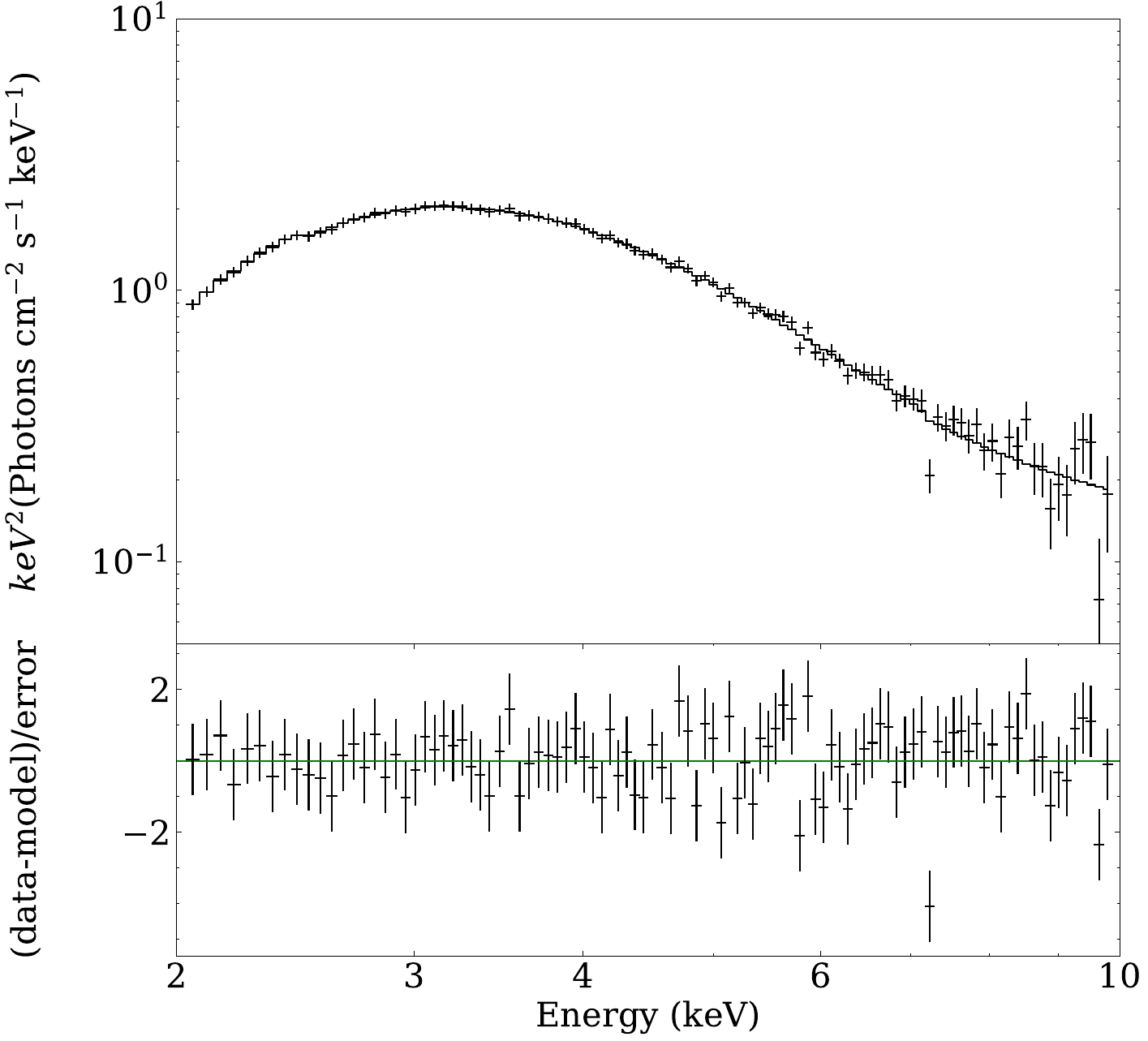}
	\caption{The Spectral fittings for the data born out of NICER observations of  Swift J151857.0--572147.}
	\label{nicer}
\end{figure}

\begin{figure*}
        \centering
%       \hspace{-1.1cm}
%       \flushleft
        \includegraphics[angle=0,scale=0.7]{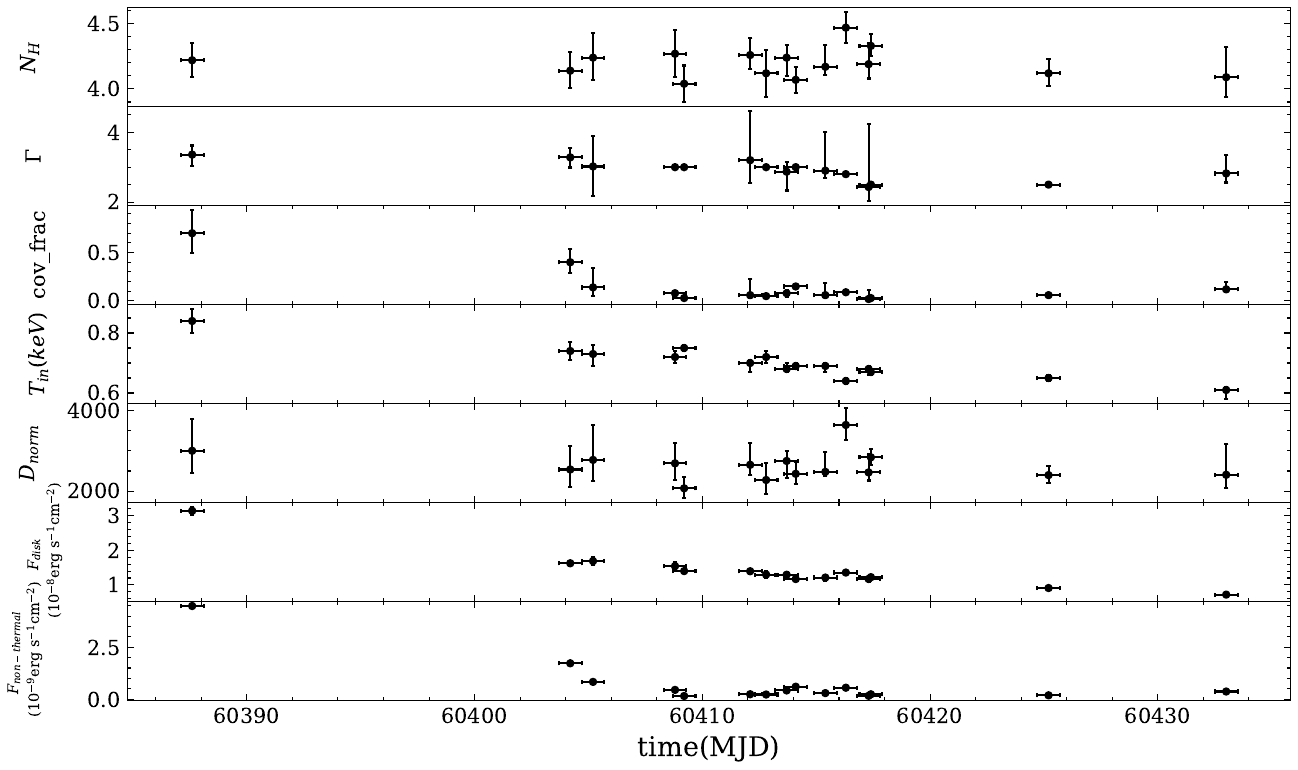}
    %\vspace{-0.3cm}
        \caption{Evolution of the spectral parameters of Swift J151857.0--572147 in 2024 from model M2: $N_{H}$ is the column density, $\Gamma$  the low-energy power-law photon index, $T_{\rm in}$ the temperature of the inner disk,  Cov\_frac the coverage factor,  and $r_{\rm in}$ the inner radius of the disk.  $F_{\rm disk}$ and $F_{\rm non-thermal}$ are disk flux and non-thermal flux, respectively.}
        
        \label{thdisk}
\end{figure*}

\begin{table}[htbp]
    %\centering
    \raggedright
    \renewcommand{\arraystretch}{1.5}
     \setlength{\tabcolsep}{3pt}
		\caption{The results of spectral fitting the Insight-HXMT+NuSTAR and NICER+NuSTAR data for Model M1.}
        \Huge
		\resizebox{8cm}{!}{
		\begin{threeparttable}
	\begin{tabular}{ccccc}
		%\begin{tabular}{|c|m{2cm}|}
		  \hline
		   \hline
        Model & Parameter &91001311002+HXMT&NICER+91001311004&
       \\ \hline
tbabs& $N_{\rm H}[10^{22} \rm cm^{-2}]$&$3.32^{+0.02}_{-0.01}$&$3.37^{+0.13}_{-0.10}$\\
\hline
diskbb&$T_{\rm in}$&$0.90_{-0.01}^{+0.01}$&$0.90_{-0.01}^{+0.01}$ \\
&norm&$1320.5_{-36.2}^{+17.1}$&$1615.5_{-113.3}^{+64.4}$\\
\hline
relxill&$a$&$0.86_{-0.14}^{+0.12}$&$0.82_{-0.22}^{+0.13}$\\
&$i$ [\textdegree]&$20.2_{-0.8}^{+0.5}$&$21.9_{-3.5}^{+4.5}$\\
&$\Gamma$&$2.56_{-0.01}^{+0.01}$&$2.63_{-0.01}^{+0.01}$\\
&$R_{\rm in}$&$-1^{*}$&$-1{*}$\\
&$R_{\rm out}$&$400^{*}$&$400^{*}$\\
&$R_{\rm br}$&$15^{*}$&$15^{*}$\\
&index1&$3^{*}$&$3^{*}$\\
&index2&$3^{*}$&$3^{*}$\\
&logxi&$4.38_{-0.06}^{+0.10}$&$4.02_{-0.21}^{+0.11}$\\
&$A_{fe}$&$5.02_{-0.23}^{+0.13}$&$3.87_{-1.53}^{+0.87}$\\
&$E_{\rm cut}$&$111.1_{-3.3}^{+7.3}$&$122.2_{-12.8}^{+21.7}$\\
&$\rm fefl\_frac$&$0.16_{-0.01}^{+0.02}$&$0.21_{-0.03}^{+0.04}$\\
&norm[$10^{-2}$]&$7.67_{-0.11}^{+0.12}$&$6.39_{-0.14}^{+0.17}$\\
\hline
constant&con[NICER]&&$1^*$\\
&con[NFPMA]&$1^*$&$1.61_{-0.06}^{+0.06}$\\
&con[NFPMB]&$0.97_{-0.01}^{+0.01}$&$1.57_{-0.06}^{+0.06}$\\
&con[ME]&$0.94_{-0.01}^{+0.01}$\\
&con[HE]&$0.82_{-0.04}^{+0.11}$\\

\hline
&$\chi^2$/(d.o.f.)&1.06&1.01\\
\hline
	\label{parameter} &
    \end{tabular}
\begin{tablenotes}[para,flushleft] 
        \item Note:\\
        **These parameters are fixed during the fitting.
     \end{tablenotes} 
     
\end{threeparttable}} 

\end{table}

\begin{table}[htbp]
    %\centering
    \raggedright
    \renewcommand{\arraystretch}{1.5}
     \setlength{\tabcolsep}{3pt}
		\caption{The results of spectral fitting the Insight-HXMT+NuSTAR and NICER+NuSTAR data for Model M1*.}
        \Huge
		\resizebox{8cm}{!}{
		\begin{threeparttable}
	\begin{tabular}{ccccc}
		%\begin{tabular}{|c|m{2cm}|}
		  \hline
		   \hline
        Model & Parameter &91001311002+HXMT&NICER+91001311004&
       \\ \hline
tbabs& $N_{\rm H}[10^{22} \rm cm^{-2}]$&$2.98^{+0.12}_{-0.09}$&$4.44^{+0.03}_{-0.01}$\\
\hline
diskbb&$T_{\rm in}$&$0.95_{-0.01}^{+0.02}$&$0.84_{-0.01}^{+0.01}$ \\
&norm&$1000.2_{-82.3}^{+53.5}$&$2242.6_{-100.9}^{+39.7}$\\
\hline
relxill&$a$&$0.87_{-0.33}^{+0.11}$&$0.86_{-0.23}^{+0.06}$\\
&$i$ [\textdegree]&$20.4_{-2.7}^{+3.1}$&$22.3_{-2.7}^{+3.0}$\\
&$\Gamma$&$2.54_{-0.02}^{+0.02}$&$2.59_{-0.01}^{+0.02}$\\
&$R_{\rm in}$&$-1^{*}$&$-1{*}$\\
&$R_{\rm out}$&$400^{*}$&$400^{*}$\\
&$R_{\rm br}$&$15^{*}$&$15^{*}$\\
&index1&$3^{*}$&$3^{*}$\\
&index2&$3^{*}$&$3^{*}$\\
&logxi&$2.30_{-0.27}^{+0.03}$&$3.30_{-0.02}^{+0.16}$\\
&$logN[cm^{-3}]$&$19.00_{-0.01}^{+0.03}$&$19.00_{-0.14}^{+0.03}$\\
&$A_{fe}$&$4.02_{-0.23}^{+0.13}$&$3.94_{-0.153}^{+0.24}$\\
&$kT_{\rm e}[keV]$&$61.5_{-13.2}^{+15.1}$&$67.2_{-12.8}^{+21.7}$\\
&$\rm fefl\_frac$&$0.19_{-0.05}^{+0.06}$&$0.10_{-0.02}^{+0.01}$\\
&norm[$10^{-2}$]&$6.02_{-0.28}^{+0.29}$&$3.49_{-0.03}^{+0.04}$\\
\hline
constant&con[NICER]&&$1^*$\\
&con[NFPMA]&$1^*$&$1.25_{-0.05}^{+0.01}$\\
&con[NFPMB]&$0.97_{-0.01}^{+0.01}$&$1.22_{-0.05}^{+0.01}$\\
&con[ME]&$0.94_{-0.01}^{+0.01}$\\
&con[HE]&$0.82_{-0.04}^{+0.11}$\\

\hline
&$\chi^2$/(d.o.f.)&1.06&1.01\\
\hline
	\label{relxillcp} &
    \end{tabular}
\begin{tablenotes}[para,flushleft] 
        \item Note:\\
        **These parameters are fixed during the fitting.
     \end{tablenotes} 
     
\end{threeparttable}} 

\end{table}

As shown in Table \ref{observation},  there are two sets of simultaneous joint observations from Insight-HXMT, NICER, and NuSTAR. One set is from Insight-HXMT and NuSTAR at the time around MJD 60386 and another set is from NICER and NuSTAR at the time around MJD 60387. Accordingly, joint spectral fittings are carried out for each of the observational sets. 

We account for interstellar absorption with the {\tt tbabs} model \citep{2000Wilms}, where photoelectric cross sections were given in \cite{1996Verner}. 
We fit the multi-temperature blackbody component of the accretion disk with  {\tt diskbb} \citep{1984Mitsuda}.
{\tt Cutoffpl} is used to fit non-thermal emission.
{\tt Constant} is introduced to balance the calibration discrepancies between different telescopes.  In order to take into account the differences between the different spectral indices of NICER and NuSTAR, we allow the temperature of the disk and spectral index to vary.
At this step, We find a poor-fitting result with 
$\chi^2$/(d.o.f)=1.34, and a significant reflection component in the residuals (Figure \ref{re}).
Therefore, we replace {\tt cutoffpl} with the {\tt relxill} V2.3 model to fit the reflectance component of the spectrum \citep{2016Dauser}, which results in a much improved fit, $\chi^2$/(d.o.f) = 1.06 (Figure \ref{nuhx}).
Therefore our fitting model M1 is: {\tt constant*tbabs*(diskbb+relxill)}.  We fix $R_{\rm in}$ at -1 and the disk at ISCO. The break radius and the outer radius of the disk are fixed at 15 and 400, respectively, and both emission indices are fixed at 3. 
The spectral parameters are shown in Tabel \ref{parameter}. The BH spin and disk inclination averaged over the two observational sets are   $0.84_{-0.26}^{+0.17}$ and $21.1_{-3.6}^{+4.5}$ degree, respectively.
The reflection fraction is relatively small and is defined in the frame of the primary source as the ratio of intensity emitted towards the disk compared to escaping to infinity. This indicates that the corona is less illuminated by the disk. The normalization of {\tt relxill} is also smaller, indicating a lower percentage of reflected fluxes.  The ionization of the accretion disk and the iron abundance of the material in the accretion disk are also higher which is related to the lower electron number density ($n_{\rm e} = 10^{15}cm^{-3}$) of the {\tt relxill}.
This relatively high metal abundance has also appeared in other BH X-ray binaries such as MAXI J1820+070, GX 339--4 and Cyg X--1 \citep{2015ApJ...808..122F,  2015ApJ...808....9P,2021NatCo..12.1025Y}.
We replace {\tt relxill} with the reflection model {\tt relxillcp} for higher density disks, so our model M1* is: constant*tbabs(diskbb+relxillcp), The parameters are shown in Table \ref {relxillcp}. The BH spin and disk inclination averaged over the two observational sets are   $0.87_{-0.20}^{+0.06}$ and $21.4_{-1.9}^{+2.2}$ degree, respectively, which is consistent with the results obtained for M1. Both the ionization parameters and the iron abundance have also decreased, but the iron abundance is still high because the density of the disk is $n_{\rm e} = 10^{15}cm^{-3}$, which is lower than the typical values in the standard thin disk model \citep{1973A&A....24..337S}.

Since the NICER observations are in the soft state of the outburst,  we fit the NICER spectrum by removing the {\tt relxill} in M1.  We find that the fit is poor, with $\chi^2$/(d.o.f)=1.41, due to the existence of a significant Compton component in the residuals. By introducing an additional {\tt thcomp} model to account for the Compton component,  the fit is largely improved with $\chi^2$/(d.o.f)=0.85  (Figure \ref{nicer}).
So our model M2 is {\tt tbabs*thcomp*diskbb} and the fluxes of the disk and non-thermal in the 0.001--100 keV are estimated with {\tt cflux}. 
The evolution of the spectral parameters is shown in Figure \ref{thdisk}, where the percentage of non-thermal emission is as low as less than 10\%, consistent with having a soft state of BHXRBs. Additionally, the normalization of {\tt diskbb} remains almost unchanged during the outburst.

\subsection{Properties of the system and the compact object}
\label{estimation}

\begin{figure}
	\centering
	\includegraphics[angle=0,scale=0.5]{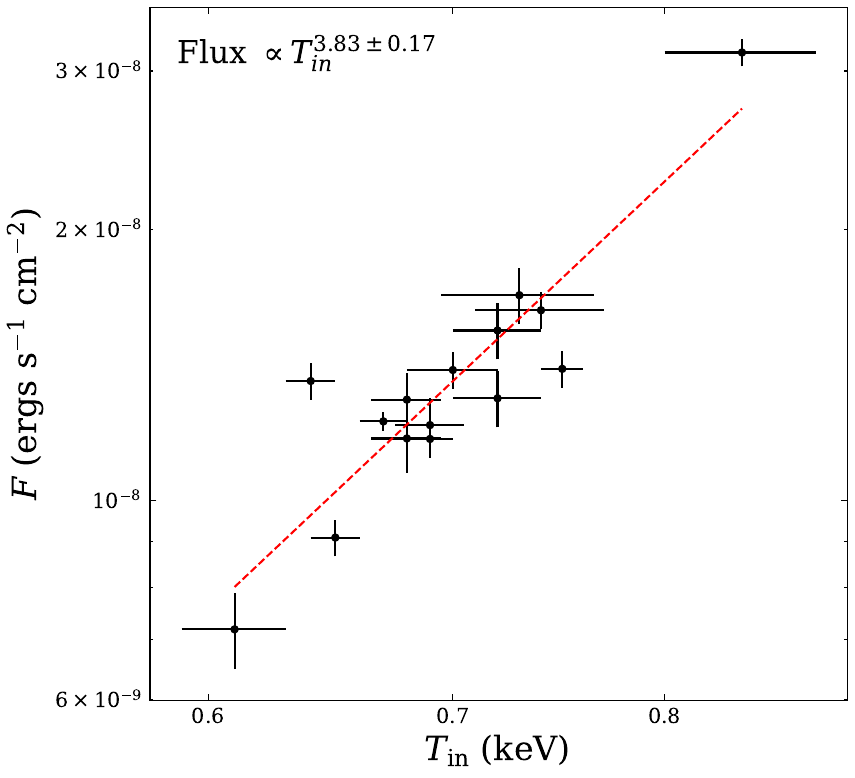}
	\caption{The disk unabsorbed flux (0.01--100 keV) versus the disk inner temperature ($T_{\rm in}$). Flux is found to vary with temperature in form  $T^{3.83\pm 0.17}_{\rm in}$. }
	\label{T4}
\end{figure}

\begin{table*}[htbp]
    \centering
		\caption{The parameters essential for distance estimation obtained by following the approach in \cite{2015Yan}, for Swift J151857.0--572147 during the 2024 outburst}
		\begin{tabular}{cccccccccccc}
		%\begin{tabular}{|c|m{2cm}|}
		  \hline
		   \hline
		  & $t_{\rm peak}$ & $F_{\rm peak}$&$\tau_{\rm decay, 10-90\%}$&$\tau_{\rm decay, 10-50\%}$&$\tau_{\rm decay, 50-90\%}$&$E$&$D$
    \\&(MJD)& (Crab) & (days)&(days)&(days)&($10^{44}$ ergs)&kpc \\ \hline
     Swift J151857.0-572147 &60377 & $1.14\pm 0.01$& $20.97\pm 0.23$& $26.23\pm 0.43$& $10.77\pm 0.33$& $1.43\pm 0.07$&$5.94\pm 4.35$ \\ 
    & & & & & & $2.18\pm 0.18$&$6.74\pm 4.94$\\
     & & & & & & $0.73\pm 0.03$&$4.72\pm 3.36$\\
     \hline 
        \label{E} &     

    \end{tabular}

\end{table*}

As shown in Figure \ref{T4},  we fit the temperature and flux of the disk obtained with M2 using the formula $F_{\rm disk}=norm*T_{\rm in}^{\alpha}$, and we find that $F_{\rm disk} \propto T_{\rm in}^{3.83\pm 0.17}$.
For the standard thin disk, the luminosity of the disk is approximately given by $L_{\rm disk}\approx 4\pi R_{\rm in}^{2}\sigma T_{\rm in}^{4}$, where $R_{\rm in}$, $T_{\rm in}$, and $\sigma$ represent the inner radius, the temperature of the disk, and the Stefan-Boltzmann constant, respectively.  So the formula between the flux and temperature of the disk we measured indicates that the inner radius of the disk is stable and the disk is in the ISCO during the HSS of the outburst. 
From {\tt diskbb} we have $\text{norm} = \left(\frac{\text{Rin}}{\text{D}_{10}}\right)^2 \cos\theta$, where $R_{\rm in}$ is the inner disk radius, $D_{\rm 10}$ is the distance of the source in units of 10 kpc, and $\theta$ is the inclination angle.  With an inclination angle of $21.1_{-3.6}^{+4.5}$ degree as aforementioned in the spectral analysis, the inner disk radius can be deduced as  $R_{\rm in}=R_{\rm ISCO}=53.2_{-1.08}^{+1.89} D_{\rm 10}$ km. However, the previous rough estimation of the distance 
$4.48^{+0.67}_{-0.47}-15.64^{+0.77}_{-0.60}$ kpc prevents from further constraining the $R_{\rm in}$. 

For the distance measurements, \cite{2015Yan} have statistically investigated the properties of 110 bright X-ray outbursts out of 36 low-mass X-ray binary transients observed by the All-Sky Monitor on RXTE from 1996--2011. They derived a number of outburst characteristics, including peak X-ray luminosity, change rate of luminosity on a daily timescale, e-folding rise and decay timescales, outburst duration, and total radiated energy. 
They built the relationships between the total radiated energy and the peak luminosity as ${\rm log} E=(-13.26\pm6.53)+(1.52\pm0.17) \times {\rm log} L_{\rm peak}$, and the relationship between the total radiated energy and the e-folding rise and decay timescales as ${\rm log} E=(43.46\pm0.10)+(0.68\pm0.11) \times {\rm log} \tau_{\rm rise,10-90\%}$, ${\rm log} E=(43.72\pm0.10)+(0.49\pm0.13) \times {\rm log} \tau_{\rm rise,10-50\%}$, ${\rm log} E=(43.48\pm0.12)+(0.60\pm0.12) \times {\rm log} \tau_{\rm rise,50-90\%}$, ${\rm log} E=(42.65\pm0.12)+(1.14\pm0.10) \times {\rm log} \tau_{\rm decay,10-90\%}$,
 ${\rm log} E=(42.65\pm0.12)+(1.00\pm0.11) \times {\rm log} \tau_{\rm decay,10-50\%}$, and 
 ${\rm log} E=(43.09\pm0.13)+(0.75\pm0.10) \times {\rm log} \tau_{\rm decay,50-90\%}$.
For Swift J151857.0--572147, we choose $F_{90\%}$, $F_{50\%}$, and $F_{10\%}$ from the light curve where the fluxes are about 90\%, 50\%, and 10\% of the peak flux (red dashed lines in Figure \ref{lcurve}). The three e-folding decay timescales are then calculated using the formula $\tau=\Delta t/\rm {ln} A$ \citep{2015Yan} and the corresponding three total radiated energies are obtained from the relationship between the total radiated energy and the e-folding decay timescales (Table \ref{E}).
The peak flux of Swift J151857.0--572147 is about 1.41 Crab, corresponding to a flux of about $2.6 \times10^{-8} { \rm erg/cm^{2}/s}$. 
As shown in Table \ref{E}, we can estimate the three distances based on the three total radiated energies obtained.
Accordingly, the average distance can be deduced as  $5.8\pm 2.5$ kpc. With this distance we have  $R_{\rm in}=30.8_{-13.3}^{+13.3}$ km.
Since the inner disk shall hold at $R_{\rm in}=2.6_{-0.04}^{+0.04}R_{\rm g}$  for a BH spin of $0.84_{-0.26}^{+0.17}$, the mass of the BH harbored in Swift J151857.0--572147 can be finally inferred as  $8.07_{-3.69}^{+3.70} M_\odot$.

\section{discussion and conclusion}
\label{diss}

We have carried out a spectral analysis of the Swift J151857.0--572147 outburst using data from NICER, Insight-HXMT and NuSTAR.
Almost the entire fundamental parameters (e.g. the BH spin, BH mass, disk inclination, and source distance, etc.) of this newly discovered  XRB system are largely uncovered for the first time.

As a newly discovered BH low-mass X-ray binary system, the fundamental parameters of Swift J151857.0--572147 are largely unknown despite a few reports claiming some rough estimations. To simultaneously constrain the spin, inclination, and mass solely with {\tt kerrbb} is usually challenging and heavily influenced by whether the distance parameter is properly set. BH spin and disk inclination can be measured in the absence of the knowledge of the BH mass and the source distance, in an approach by fitting the disk reflection feature and the broadened iron line. Fortunately, the joint Insight-HXMT, NICER and NuSTAR observations of Swift J151857.0--572147 provide data with high quality and broad energy coverage, essential for disentangling in the spectrum the reflection component. As a result, the BH spin and the disk inclination Swift J151857.0--572147 are precisely refined as $0.84_{-0.26}^{+0.17}$ and $21.1_{-3.6}^{+4.5}$ degree, respectively.  As shown in Table \ref{parameter}, adoption of the flavor of  relxill can well account for the reflection component. We notice that an alternative flavor of {\tt relxillcp} can also fit the spectrum and have little influence on the measurements of BH spin and inclination angle. An unusually high abundance is required from the measured $A_{\rm fe}$, which is also reported in outbursts of some other BH XRB systems of e.g. MAXI J1820+070, GX 339--4 and Cyg X--1 \citep{2015ApJ...808..122F,  2015ApJ...808....9P,2021NatCo..12.1025Y}. We try a version of {\tt relxill} with a high disk density ({\tt relxllcp}) and find the strange abundance value is an artifact of assuming a low density. Another uncertainty comes from the manner we eliminate the calibration difference by allowing the parameters to vary across instruments of NICER and NuSTAR. We try as well instead with constant and find that, while the fitting gets worse the BH spin and disk inclination can be inferred with fewer changes.  Hence we think the BH spin and inclination angle are measured more or less robust.

Swift J151857.0--572147 was observed to have an outburst on March 3, 2024 (MJD 60372), and enter a soft state on March 10, 2024 (MJD 60379) \citep{2024Kenneal,2024Del}. The peak flux was caught by Insight-HXMT and the soft state was monitored by NICER. Our spectral analysis of the  NICER data reveals that in the soft state, the disk flux is proportional to $T_{\rm in}^{3.83\pm0.17}$, indicating an inner disk staying around ISCO.
Since with a BH spin of $0.84_{-0.26}^{+0.17}$ and a disk normalization of $2628.9_{-90.2}^{+132.9}$ for Swift J151857.0--572147 we have $R_{\rm in}=R_{\rm ISCO}=2.6_{-0.4}^{+0.4}R_{\rm g}=53.2_{-1.08}^{+1.59} D_{\rm 10}$ km, the BH mass can in principle be directly inferred once the source distance is known. A distance of 
$4.48^{+0.67}_{-0.47}-15.64^{+0.77}_{-0.60}$ kpc claimed previously in \cite{2024B} highlights the huge uncertainty in distance measurement. As a result, BH mass can fall in a range of $6.24^{+1.35}_{-1.17}-21.77^{+3.58}_{-3.48} M_\odot$, with a notable range of uncertainty. Obviously, refined source distance would be essential for further probing the mass of the BH harbored in Swift J151857.0--572147.

\cite{2015Yan} conducted a statistical analysis of 110 bright X-ray outbursts from 36 low-mass X-ray binary transients observed by the All-Sky Monitor on RXTE between 1996 and 2011. They measured several outburst characteristics, including peak X-ray luminosity, daily rate of change of luminosity, e-folding rise and decay timescales, outburst duration, and total radiated energy. 
Since the peak X-ray luminosity is a function of disk mass, which is expected from the disk instability
model (DIM) \citep{1998MNRAS.293L..42K,2001NewAR..45..449L}, disk mass plays a major role in determining outburst properties \citep{2007ApJ...667.1043Y}.
If one can assume as an approximation that the radiation efficiency is constant, the total radiant energy E should correspond to the disk mass accreted during the outburst. Therefore, they found a positive correlation between the total radiative energy and the peak X-ray luminosity, as well as a positive correlation between the total radiative energy and the time scale of e-folding rise or decay in the outburst.
With these relations, distances were estimated for SLX 1746--331, SWIFT J1539.2--6227, SWIFT J1842.5--1124, XTE J1755--324 and XTE J2012+381.
By taking the approach in \cite{2015Yan},  and the outburst profile monitored jointly by Insight-HXMT and NICER at 2--12 keV, a distance of $5.8\pm 2.5$ kpc can be assigned to Swift J151857.0--572147.  Finally, the joint diagnostic upon the systematic properties of Swift J151857.0--572147 ends with a BH mass measured as  $8.07_{-3.69}^{+3.70} M_\odot$.

We would like to note that, such an estimation of BH mass may suffer from a few aspects in measuring the inner disk radius with a simplistic version of {\tt diskbb}. (1) The {\tt diskbb} model does not include a zero-torque inner boundary condition. Using a model that includes this important physical effect tends to reduce radii by a factor of about 2.2 (See e.g., \cite{2005ApJ...618..832Z}), which may reduce the BH mass from $\sim$ $8.07_{-3.69}^{+3.70} M_\odot$ to $ 3.67_{-1.68}^{+1.68} M_\odot$.  (2) There is a color correction factor that aims to account for scattering in the disk atmosphere.  The color correction factor $f_{\rm col}$ is 1.7$\pm$0.2 and is minimally affected by mass and accretion rate, indicating a weak dependence of $f_{\rm col}$ on black hole spin $a$ \citep{1995ApJ...445..780S}. (3) Additionally, there are also relativistic corrections to the {\tt diskbb} results.
 \cite{1997Zhang} found that the uncertainty in $f_{\rm col}$ is 10\%, which will cause an error of no more than 20\% in the inner radius of the disk.    Therefore, the final conservative estimation of BH mass is $3.67\pm1.79-8.07\pm 4.20 M_\odot$. We find that, with such a BH mass and aforementioned distance, the outburst gets a peak flux of $0.09\pm 0.08-0.19\pm0.16 L_{\rm Edd}$, and the disk is expected to be described as a thin disk that is not truncated, and not too thick at its smallest radius.

As addressed in \cite{2023Draghis,2024Draghis}, BH spin has a series of impacts on our understanding of the outburst behavior for emissions in different energy bands and as well the possible originations of the BH. For the former, the inner disk can reach around the event horizon with high radiation efficiency for an extremely spinning BH.  For the latter, it seems that so far we have two populations of BH: the group from observing the BH mergers with the gravitational wave experiments has slow spin, while those observed from XRB outbursts are always highly spinning \cite{2024Draghis}. The causality of the different BH populations is not yet known and the discovery of a BH spin in Swift J151857.0--572147 enlarges the BH spin sample, which so far consists of only about 40 BHs held with XRBs, and thus helps to disentangle the models at work in understanding the possible formations of these two BH populations.

%% Also note that the akcnowlodgment environment does not support long amounts of text. If you have a lot of people and institutions to acknowledge, do not use this command. Instead, create a new \section{Acknowledgments}.
\begin{acknowledgments}
This work is supported by the National Key R\&D Program of China (2021YFA0718500), the National Natural Science Foundation of China under grants No. 12333007, U1838202,  U2038101, U1938103, 12273030, U1938107, 12027803 and 12173103.
This work made use of data and software from the Insight-HXMT mission, a project funded by the China National Space Administration (CNSA) and the Chinese Academy of Sciences(CAS). This work was partially supported by the International Partnership Program of the Chinese Academy of Sciences (Grant No.113111KYSB20190020).
This research has made use of software provided by data obtained from the High Energy Astrophysics Science Archive Research Center (HEASARC), provided by NASA’s Goddard Space Flight Center.
L. D. Kong is grateful for the financial support provided by the Sino-German (CSC-DAAD) Postdoc Scholarship Program (91839752).
\end{acknowledgments}





\bibliography{sample631}{}

\begin{thebibliography}{}
\expandafter\ifx\csname natexlab\endcsname\relax\def\natexlab#1{#1}\fi
\providecommand{\url}[1]{\href{#1}{#1}}
\providecommand{\dodoi}[1]{doi:~\href{http://doi.org/#1}{\nolinkurl{#1}}}
\providecommand{\doeprint}[1]{\href{http://ascl.net/#1}{\nolinkurl{http://ascl.net/#1}}}
\providecommand{\doarXiv}[1]{\href{https://arxiv.org/abs/#1}{\nolinkurl{https://arxiv.org/abs/#1}}}

\bibitem[{{Belloni} {et~al.}(2005){Belloni}, {Homan}, {Casella}, {van der Klis}, {Nespoli}, {Lewin}, {Miller}, \& {M{\'e}ndez}}]{2005Belloni}
{Belloni}, T., {Homan}, J., {Casella}, P., {et~al.} 2005, \aap, 440, 207, \dodoi{10.1051/0004-6361:20042457}

\bibitem[{{Blandford} \& {Znajek}(1977)}]{1977MNRAS.179..433B}
{Blandford}, R.~D., \& {Znajek}, R.~L. 1977, \mnras, 179, 433, \dodoi{10.1093/mnras/179.3.433}

\bibitem[{{Bodensteiner} {et~al.}(2022){Bodensteiner}, {Heida}, {Abdul-Masih}, {Baade}, {Banyard}, {Bowman}, {Fabry}, {Frost}, {Mahy}, {Marchant}, {M{\'e}rand}, {Reggiani}, {Rivinius}, {Sana}, {Selman}, \& {Shenar}}]{2022Msngr.186....3B}
{Bodensteiner}, J., {Heida}, M., {Abdul-Masih}, M., {et~al.} 2022, The Messenger, 186, 3, \dodoi{10.18727/0722-6691/5255}

\bibitem[{{Brenneman} \& {Reynolds}(2006)}]{2006Brenneman}
{Brenneman}, L.~W., \& {Reynolds}, C.~S. 2006, \apj, 652, 1028, \dodoi{10.1086/508146}

\bibitem[{{Burridge} {et~al.}(2024){Burridge}, {Miller-Jones}, {Bahramian}, {Prabu}, {Carotenuto}, {Russell}, {Cowie}, \& {Fender}}]{2024B}
{Burridge}, B.~J., {Miller-Jones}, J.~C.~A., {Bahramian}, A., {et~al.} 2024, The Astronomer's Telegram, 16538, 1

\bibitem[{{Campanelli} {et~al.}(2006){Campanelli}, {Lousto}, \& {Zlochower}}]{2006PhRvD..74d1501C}
{Campanelli}, M., {Lousto}, C.~O., \& {Zlochower}, Y. 2006, \prd, 74, 041501, \dodoi{10.1103/PhysRevD.74.041501}

\bibitem[{{Cao} {et~al.}(2020){Cao}, {Jiang}, {Meng}, {Zhang}, {Luo}, {Yang}, {Zhang}, {Gu}, {Sun}, {Liu}, {Yang}, {Li}, {Tan}, {Liu}, {Du}, {Lu}, {Xu}, {Guan}, {Zhang}, {Wang}, {Li}, {Zhang}, {Wen}, {Qu}, {Song}, {Li}, {Ge}, {Zhou}, {Xiong}, {Zhang}, {Zhang}, {Cheng}, {Zhang}, {Li}, {Liang}, {Gao}, {Yang}, {Liu}, {Liu}, {Yang}, \& {Zhang}}]{2020Cao}
{Cao}, X., {Jiang}, W., {Meng}, B., {et~al.} 2020, Science China Physics, Mechanics, and Astronomy, 63, 249504, \dodoi{10.1007/s11433-019-1506-1}

\bibitem[{{Carotenuto} \& {Russell}(2024)}]{2024Car}
{Carotenuto}, F., \& {Russell}, T.~D. 2024, The Astronomer's Telegram, 16518, 1

\bibitem[{{Chatterjee} {et~al.}(2024){Chatterjee}, {Pujitha Suribhatla}, {Mondal}, \& {Singh}}]{2024arXiv240617629C}
{Chatterjee}, K., {Pujitha Suribhatla}, S., {Mondal}, S., \& {Singh}, C.~B. 2024, arXiv e-prints, arXiv:2406.17629, \dodoi{10.48550/arXiv.2406.17629}

\bibitem[{{Chen} {et~al.}(2020){Chen}, {Cui}, {Li}, {Wang}, {Xu}, {Lu}, {Wang}, {Chen}, {Han}, {Hu}, {Zhang}, {Huo}, {Yang}, {Li}, {Lu}, {Zhang}, {Li}, {Zhang}, {Xiong}, {Zhang}, {Xue}, {Zhao}, {Zhu}, {Zhu}, {Liu}, {Yang}, \& {Zhang}}]{2020Chen}
{Chen}, Y., {Cui}, W., {Li}, W., {et~al.} 2020, Science China Physics, Mechanics, and Astronomy, 63, 249505, \dodoi{10.1007/s11433-019-1469-5}

\bibitem[{{Cowie} {et~al.}(2024){Cowie}, {Carotenuto}, {Fender}, {Heywood}, {Hughes}, {Sivakoff}, \& {X-KAT Collaboration}}]{2024ATel16503....1C}
{Cowie}, F.~J., {Carotenuto}, F., {Fender}, R.~P., {et~al.} 2024, The Astronomer's Telegram, 16503, 1

\bibitem[{{Dauser} {et~al.}(2016){Dauser}, {Garc{\'\i}a}, \& {Wilms}}]{2016Dauser}
{Dauser}, T., {Garc{\'\i}a}, J., \& {Wilms}, J. 2016, Astronomische Nachrichten, 337, 362, \dodoi{10.1002/asna.201612314}

\bibitem[{{Del Santo} {et~al.}(2024){Del Santo}, {Russell}, {Marino}, \& {Motta}}]{2024Del}
{Del Santo}, M., {Russell}, T.~D., {Marino}, A., \& {Motta}, S. 2024, The Astronomer's Telegram, 16519, 1

\bibitem[{{Draghis} {et~al.}(2024){Draghis}, {Miller}, {Costantini}, {Gallo}, {Reynolds}, {Tomsick}, \& {Zoghbi}}]{2024Draghis}
{Draghis}, P.~A., {Miller}, J.~M., {Costantini}, E., {et~al.} 2024, \apj, 969, 40, \dodoi{10.3847/1538-4357/ad43ea}

\bibitem[{{Draghis} {et~al.}(2023){Draghis}, {Miller}, {Zoghbi}, {Reynolds}, {Costantini}, {Gallo}, \& {Tomsick}}]{2023Draghis}
{Draghis}, P.~A., {Miller}, J.~M., {Zoghbi}, A., {et~al.} 2023, \apj, 946, 19, \dodoi{10.3847/1538-4357/acafe7}

\bibitem[{{Esin} {et~al.}(1997){Esin}, {McClintock}, \& {Narayan}}]{1997Esin}
{Esin}, A.~A., {McClintock}, J.~E., \& {Narayan}, R. 1997, \apj, 489, 865, \dodoi{10.1086/304829}

\bibitem[{{Fender} {et~al.}(2004){Fender}, {Belloni}, \& {Gallo}}]{2004Fender}
{Fender}, R.~P., {Belloni}, T.~M., \& {Gallo}, E. 2004, \mnras, 355, 1105, \dodoi{10.1111/j.1365-2966.2004.08384.x}

\bibitem[{{F{\"u}rst} {et~al.}(2015){F{\"u}rst}, {Nowak}, {Tomsick}, {Miller}, {Corbel}, {Bachetti}, {Boggs}, {Christensen}, {Craig}, {Fabian}, {Gandhi}, {Grinberg}, {Hailey}, {Harrison}, {Kara}, {Kennea}, {Madsen}, {Pottschmidt}, {Stern}, {Walton}, {Wilms}, \& {Zhang}}]{2015ApJ...808..122F}
{F{\"u}rst}, F., {Nowak}, M.~A., {Tomsick}, J.~A., {et~al.} 2015, \apj, 808, 122, \dodoi{10.1088/0004-637X/808/2/122}

\bibitem[{{Gendreau} {et~al.}(2016){Gendreau}, {Arzoumanian}, {Adkins}, {Albert}, {Anders}, {Aylward}, {Baker}, {Balsamo}, {Bamford}, {Benegalrao}, {Berry}, {Bhalwani}, {Black}, {Blaurock}, {Bronke}, {Brown}, {Budinoff}, {Cantwell}, {Cazeau}, {Chen}, {Clement}, {Colangelo}, {Coleman}, {Coopersmith}, {Dehaven}, {Doty}, {Egan}, {Enoto}, {Fan}, {Ferro}, {Foster}, {Galassi}, {Gallo}, {Green}, {Grosh}, {Ha}, {Hasouneh}, {Heefner}, {Hestnes}, {Hoge}, {Jacobs}, {J{\o}rgensen}, {Kaiser}, {Kellogg}, {Kenyon}, {Koenecke}, {Kozon}, {LaMarr}, {Lambertson}, {Larson}, {Lentine}, {Lewis}, {Lilly}, {Liu}, {Malonis}, {Manthripragada}, {Markwardt}, {Matonak}, {Mcginnis}, {Miller}, {Mitchell}, {Mitchell}, {Mohammed}, {Monroe}, {Montt de Garcia}, {Mul{\'e}}, {Nagao}, {Ngo}, {Norris}, {Norwood}, {Novotka}, {Okajima}, {Olsen}, {Onyeachu}, {Orosco}, {Peterson}, {Pevear}, {Pham}, {Pollard}, {Pope}, {Powers}, {Powers}, {Price}, {Prigozhin}, {Ramirez}, {Reid}, {Remillard}, {Rogstad}, {Rosecrans}, {Rowe}, {Sager}, {Sanders},
  {Savadkin}, {Saylor}, {Schaeffer}, {Schweiss}, {Semper}, {Serlemitsos}, {Shackelford}, {Soong}, {Struebel}, {Vezie}, {Villasenor}, {Winternitz}, {Wofford}, {Wright}, {Yang}, \& {Yu}}]{2016Gendreau}
{Gendreau}, K.~C., {Arzoumanian}, Z., {Adkins}, P.~W., {et~al.} 2016, in Society of Photo-Optical Instrumentation Engineers (SPIE) Conference Series, Vol. 9905, Space Telescopes and Instrumentation 2016: Ultraviolet to Gamma Ray, ed. J.-W.~A. {den Herder}, T.~{Takahashi}, \& M.~{Bautz}, 99051H, \dodoi{10.1117/12.2231304}

\bibitem[{{Gierli{\'n}ski} {et~al.}(2008){Gierli{\'n}ski}, {Done}, \& {Page}}]{2008Gierlinski}
{Gierli{\'n}ski}, M., {Done}, C., \& {Page}, K. 2008, \mnras, 388, 753, \dodoi{10.1111/j.1365-2966.2008.13431.x}

\bibitem[{{Harrison} {et~al.}(2013){Harrison}, {Craig}, {Christensen}, {Hailey}, {Zhang}, {Boggs}, {Stern}, {Cook}, {Forster}, {Giommi}, {Grefenstette}, {Kim}, {Kitaguchi}, {Koglin}, {Madsen}, {Mao}, {Miyasaka}, {Mori}, {Perri}, {Pivovaroff}, {Puccetti}, {Rana}, {Westergaard}, {Willis}, {Zoglauer}, {An}, {Bachetti}, {Barri{\`e}re}, {Bellm}, {Bhalerao}, {Brejnholt}, {Fuerst}, {Liebe}, {Markwardt}, {Nynka}, {Vogel}, {Walton}, {Wik}, {Alexander}, {Cominsky}, {Hornschemeier}, {Hornstrup}, {Kaspi}, {Madejski}, {Matt}, {Molendi}, {Smith}, {Tomsick}, {Ajello}, {Ballantyne}, {Balokovi{\'c}}, {Barret}, {Bauer}, {Blandford}, {Brandt}, {Brenneman}, {Chiang}, {Chakrabarty}, {Chenevez}, {Comastri}, {Dufour}, {Elvis}, {Fabian}, {Farrah}, {Fryer}, {Gotthelf}, {Grindlay}, {Helfand}, {Krivonos}, {Meier}, {Miller}, {Natalucci}, {Ogle}, {Ofek}, {Ptak}, {Reynolds}, {Rigby}, {Tagliaferri}, {Thorsett}, {Treister}, \& {Urry}}]{2013Harrison}
{Harrison}, F.~A., {Craig}, W.~W., {Christensen}, F.~E., {et~al.} 2013, \apj, 770, 103, \dodoi{10.1088/0004-637X/770/2/103}

\bibitem[{{Homan} {et~al.}(2001){Homan}, {Wijnands}, {van der Klis}, {Belloni}, {van Paradijs}, {Klein-Wolt}, {Fender}, \& {M{\'e}ndez}}]{2001Homan}
{Homan}, J., {Wijnands}, R., {van der Klis}, M., {et~al.} 2001, \apjs, 132, 377, \dodoi{10.1086/318954}

\bibitem[{{Kalogera} \& {Baym}(1996)}]{1996ApJ...470L..61K}
{Kalogera}, V., \& {Baym}, G. 1996, \apjl, 470, L61, \dodoi{10.1086/310296}

\bibitem[{{Kennea} {et~al.}(2024{\natexlab{a}}){Kennea}, {Lien}, {D'Elia}, {Melandri}, {Page}, \& {Siegel}}]{2024ATel16500....1K}
{Kennea}, J.~A., {Lien}, A.~Y., {D'Elia}, V., {et~al.} 2024{\natexlab{a}}, The Astronomer's Telegram, 16500, 1

\bibitem[{{Kennea} {et~al.}(2024{\natexlab{b}}){Kennea}, {Lien}, {D'Elia}, {Melandri}, {Page}, \& {Siegel}}]{2024Kenneal}
---. 2024{\natexlab{b}}, The Astronomer's Telegram, 16500, 1

\bibitem[{{King} \& {Ritter}(1998)}]{1998MNRAS.293L..42K}
{King}, A.~R., \& {Ritter}, H. 1998, \mnras, 293, L42, \dodoi{10.1046/j.1365-8711.1998.01295.x}

\bibitem[{{Lasota}(2001)}]{2001NewAR..45..449L}
{Lasota}, J.-P. 2001, \nar, 45, 449, \dodoi{10.1016/S1387-6473(01)00112-9}

\bibitem[{{Li} {et~al.}(2005){Li}, {Zimmerman}, {Narayan}, \& {McClintock}}]{2005Li}
{Li}, L.-X., {Zimmerman}, E.~R., {Narayan}, R., \& {McClintock}, J.~E. 2005, \apjs, 157, 335, \dodoi{10.1086/428089}

\bibitem[{{Liao} {et~al.}(2020){Liao}, {Zhang}, {Lu}, {Zhang}, {Li}, {Chang}, {Chen}, {Ge}, {Guo}, {Huang}, {Jin}, {Li}, {Li}, {Li}, {Liu}, {Lu}, {Nie}, {Song}, {Wang}, {You}, {Zhang}, {Zhao}, \& {Zhang}}]{2020Liao}
{Liao}, J.-Y., {Zhang}, S., {Lu}, X.-F., {et~al.} 2020, Journal of High Energy Astrophysics, 27, 14, \dodoi{10.1016/j.jheap.2020.04.002}

\bibitem[{{Liu} {et~al.}(2020){Liu}, {Zhang}, {Li}, {Lu}, {Chang}, {Li}, {Zhang}, {Jin}, {Yu}, {Zhang}, {Fu}, {Chen}, {Ji}, {Xu}, {Deng}, {Shang}, {Liu}, {Lu}, {Zhang}, {Dong}, {Li}, {Wu}, {Li}, {Wang}, {Wu}, {Zhang}, {Zhang}, {Xiong}, {Liu}, {Zhang}, {Liu}, {Yang}, \& {Zhang}}]{2020Liu}
{Liu}, C., {Zhang}, Y., {Li}, X., {et~al.} 2020, Science China Physics, Mechanics, and Astronomy, 63, 249503, \dodoi{10.1007/s11433-019-1486-x}

\bibitem[{{McClintock} {et~al.}(2006){McClintock}, {Shafee}, {Narayan}, {Remillard}, {Davis}, \& {Li}}]{2006ApJ...652..518M}
{McClintock}, J.~E., {Shafee}, R., {Narayan}, R., {et~al.} 2006, \apj, 652, 518, \dodoi{10.1086/508457}

\bibitem[{{Miller} {et~al.}(2006){Miller}, {Homan}, {Steeghs}, {Rupen}, {Hunstead}, {Wijnands}, {Charles}, \& {Fabian}}]{2006Miller}
{Miller}, J.~M., {Homan}, J., {Steeghs}, D., {et~al.} 2006, \apj, 653, 525, \dodoi{10.1086/508644}

\bibitem[{{Miller} {et~al.}(2009){Miller}, {Reynolds}, {Fabian}, {Miniutti}, \& {Gallo}}]{2009Miller}
{Miller}, J.~M., {Reynolds}, C.~S., {Fabian}, A.~C., {Miniutti}, G., \& {Gallo}, L.~C. 2009, \apj, 697, 900, \dodoi{10.1088/0004-637X/697/1/900}

\bibitem[{{Mitsuda} {et~al.}(1984){Mitsuda}, {Inoue}, {Koyama}, {Makishima}, {Matsuoka}, {Ogawara}, {Shibazaki}, {Suzuki}, {Tanaka}, \& {Hirano}}]{1984Mitsuda}
{Mitsuda}, K., {Inoue}, H., {Koyama}, K., {et~al.} 1984, \pasj, 36, 741

\bibitem[{{Mondal} {et~al.}(2024){Mondal}, {Pujitha Suribhatla}, {Chatterjee}, {Singh}, \& {Chatterjee}}]{2024M}
{Mondal}, S., {Pujitha Suribhatla}, S., {Chatterjee}, K., {Singh}, C.~B., \& {Chatterjee}, R. 2024, arXiv e-prints, arXiv:2404.09643, \dodoi{10.48550/arXiv.2404.09643}

\bibitem[{{Motta} {et~al.}(2012){Motta}, {Homan}, {Mu{\~n}oz Darias}, {Casella}, {Belloni}, {Hiemstra}, \& {M{\'e}ndez}}]{2012Motta}
{Motta}, S., {Homan}, J., {Mu{\~n}oz Darias}, T., {et~al.} 2012, \mnras, 427, 595, \dodoi{10.1111/j.1365-2966.2012.22037.x}

\bibitem[{{Parker} {et~al.}(2015){Parker}, {Tomsick}, {Miller}, {Yamaoka}, {Lohfink}, {Nowak}, {Fabian}, {Alston}, {Boggs}, {Christensen}, {Craig}, {F{\"u}rst}, {Gandhi}, {Grefenstette}, {Grinberg}, {Hailey}, {Harrison}, {Kara}, {King}, {Stern}, {Walton}, {Wilms}, \& {Zhang}}]{2015ApJ...808....9P}
{Parker}, M.~L., {Tomsick}, J.~A., {Miller}, J.~M., {et~al.} 2015, \apj, 808, 9, \dodoi{10.1088/0004-637X/808/1/9}

\bibitem[{{Peng} {et~al.}(2023){Peng}, {Zhang}, {Wang}, {Zhang}, {Kong}, {Chen}, {Shui}, {Ji}, {Qu}, {Tao}, {Ge}, {Ma}, {Chang}, {Li}, {Li}, {Yu}, {Yan}, {Zhang}, {Xiao}, \& {Zhao}}]{2023P}
{Peng}, J.-Q., {Zhang}, S., {Wang}, P.-J., {et~al.} 2023, \apj, 955, 96, \dodoi{10.3847/1538-4357/acf461}

\bibitem[{{Peng} {et~al.}(2024){Peng}, {Zhang}, {Shui}, {Zhang}, {Chen}, {Kong}, {Yu}, {Ji}, {Wang}, {Ge}, {Qu}, {Tao}, {Chang}, {Li}, {Li}, \& {Yan}}]{2024ApJ...965L..22P}
{Peng}, J.-Q., {Zhang}, S., {Shui}, Q.-C., {et~al.} 2024, \apjl, 965, L22, \dodoi{10.3847/2041-8213/ad3640}

\bibitem[{{Reynolds} \& {Miller}(2013)}]{2013Reynolds}
{Reynolds}, M.~T., \& {Miller}, J.~M. 2013, \apj, 769, 16, \dodoi{10.1088/0004-637X/769/1/16}

\bibitem[{{Shakura} \& {Sunyaev}(1973)}]{1973A&A....24..337S}
{Shakura}, N.~I., \& {Sunyaev}, R.~A. 1973, \aap, 24, 337

\bibitem[{{Shimura} \& {Takahara}(1995)}]{1995ApJ...445..780S}
{Shimura}, T., \& {Takahara}, F. 1995, \apj, 445, 780, \dodoi{10.1086/175740}

\bibitem[{{Steiner} {et~al.}(2011){Steiner}, {Reis}, {McClintock}, {Narayan}, {Remillard}, {Orosz}, {Gou}, {Fabian}, \& {Torres}}]{2011MNRAS.416..941S}
{Steiner}, J.~F., {Reis}, R.~C., {McClintock}, J.~E., {et~al.} 2011, \mnras, 416, 941, \dodoi{10.1111/j.1365-2966.2011.19089.x}

\bibitem[{{Tetarenko} {et~al.}(2016){Tetarenko}, {Sivakoff}, {Heinke}, \& {Gladstone}}]{2016ApJS..222...15T}
{Tetarenko}, B.~E., {Sivakoff}, G.~R., {Heinke}, C.~O., \& {Gladstone}, J.~C. 2016, \apjs, 222, 15, \dodoi{10.3847/0067-0049/222/2/15}

\bibitem[{{Verner} {et~al.}(1996){Verner}, {Ferland}, {Korista}, \& {Yakovlev}}]{1996Verner}
{Verner}, D.~A., {Ferland}, G.~J., {Korista}, K.~T., \& {Yakovlev}, D.~G. 1996, \apj, 465, 487, \dodoi{10.1086/177435}

\bibitem[{{Wilms} {et~al.}(2000){Wilms}, {Allen}, \& {McCray}}]{2000Wilms}
{Wilms}, J., {Allen}, A., \& {McCray}, R. 2000, \apj, 542, 914, \dodoi{10.1086/317016}

\bibitem[{{Woosley}(1993)}]{1993ApJ...405..273W}
{Woosley}, S.~E. 1993, \apj, 405, 273, \dodoi{10.1086/172359}

\bibitem[{{Yan} \& {Yu}(2015)}]{2015Yan}
{Yan}, Z., \& {Yu}, W. 2015, \apj, 805, 87, \dodoi{10.1088/0004-637X/805/2/87}

\bibitem[{{You} {et~al.}(2021){You}, {Tuo}, {Li}, {Wang}, {Zhang}, {Zhang}, {Ge}, {Luo}, {Liu}, {Yuan}, {Dai}, {Liu}, {Qiao}, {Jin}, {Liu}, {Czerny}, {Wu}, {Bu}, {Cai}, {Cao}, {Chang}, {Chen}, {Chen}, {Chen}, {Chen}, {Chen}, {Chen}, {Cui}, {Cui}, {Deng}, {Dong}, {Du}, {Fu}, {Gao}, {Gao}, {Gao}, {Gu}, {Guan}, {Guo}, {Han}, {Huang}, {Huo}, {Jia}, {Jiang}, {Jiang}, {Jin}, {Jin}, {Kong}, {Li}, {Li}, {Li}, {Li}, {Li}, {Li}, {Li}, {Li}, {Li}, {Li}, {Li}, {Liang}, {Liao}, {Liu}, {Liu}, {Liu}, {Liu}, {Liu}, {Lu}, {Lu}, {Lu}, {Luo}, {Luo}, {Ma}, {Meng}, {Nang}, {Nie}, {Ou}, {Qu}, {Sai}, {Shang}, {Song}, {Song}, {Sun}, {Tan}, {Tao}, {Wang}, {Wang}, {Wang}, {Wang}, {Wang}, {Wang}, {Wen}, {Wu}, {Wu}, {Wu}, {Xiao}, {Xiao}, {Xiong}, {Xu}, {Yang}, {Yang}, {Yang}, {Yi}, {Yin}, {You}, {Zhang}, {Zhang}, {Zhang}, {Zhang}, {Zhang}, {Zhang}, {Zhang}, {Zhang}, {Zhang}, {Zhang}, {Zhang}, {Zhang}, {Zhang}, {Zhang}, {Zhang}, {Zhao}, {Zhao}, {Zheng}, {Zhou}, {Zhou}, {Zhu}, \& {Zhu}}]{2021NatCo..12.1025Y}
{You}, B., {Tuo}, Y., {Li}, C., {et~al.} 2021, Nature Communications, 12, 1025, \dodoi{10.1038/s41467-021-21169-5}

\bibitem[{{Yu} \& {Dolence}(2007)}]{2007ApJ...667.1043Y}
{Yu}, W., \& {Dolence}, J. 2007, \apj, 667, 1043, \dodoi{10.1086/521011}

\bibitem[{{Zhang} {et~al.}(2014){Zhang}, {Lu}, {Zhang}, \& {Li}}]{2014Zhang}
{Zhang}, S., {Lu}, F.~J., {Zhang}, S.~N., \& {Li}, T.~P. 2014, in Society of Photo-Optical Instrumentation Engineers (SPIE) Conference Series, Vol. 9144, Space Telescopes and Instrumentation 2014: Ultraviolet to Gamma Ray, ed. T.~{Takahashi}, J.-W.~A. {den Herder}, \& M.~{Bautz}, 914421, \dodoi{10.1117/12.2054144}

\bibitem[{{Zhang} {et~al.}(2018){Zhang}, {Zhang}, {Lu}, {Li}, {Song}, {Xu}, {Wang}, {Qu}, {Liu}, {Chen}, {Cao}, {Zhang}, {Xiong}, {Ge}, {Chen}, {Liao}, {Nie}, {Zhao}, {Jia}, {Li}, {Guan}, {Li}, {Zhang}, {Jin}, {Wang}, {Zheng}, {Ma}, {Tao}, \& {Huang}}]{2018Zhang}
{Zhang}, S., {Zhang}, S.~N., {Lu}, F.~J., {et~al.} 2018, in Society of Photo-Optical Instrumentation Engineers (SPIE) Conference Series, Vol. 10699, Space Telescopes and Instrumentation 2018: Ultraviolet to Gamma Ray, ed. J.-W.~A. {den Herder}, S.~{Nikzad}, \& K.~{Nakazawa}, 106991U, \dodoi{10.1117/12.2311835}

\bibitem[{{Zhang} {et~al.}(1997){Zhang}, {Cui}, \& {Chen}}]{1997Zhang}
{Zhang}, S.~N., {Cui}, W., \& {Chen}, W. 1997, \apjl, 482, L155, \dodoi{10.1086/310705}

\bibitem[{{Zhang} {et~al.}(2020){Zhang}, {Li}, {Lu}, {Song}, {Xu}, {Liu}, {Chen}, {Cao}, {Bu}, {Chang}, {Chen}, {Chen}, {Chen}, {Chen}, {Chen}, {Cui}, {Cui}, {Deng}, {Dong}, {Du}, {Fu}, {Gao}, {Gao}, {Gao}, {Ge}, {Gu}, {Guan}, {Gungor}, {Guo}, {Han}, {Hu}, {Huang}, {Huo}, {Jia}, {Jiang}, {Jiang}, {Jin}, {Jin}, {Li}, {Li}, {Li}, {Li}, {Li}, {Li}, {Li}, {Li}, {Li}, {Li}, {Li}, {Liang}, {Liao}, {Liu}, {Liu}, {Liu}, {Liu}, {Liu}, {Liu}, {Lu}, {Lu}, {Luo}, {Ma}, {Meng}, {Nang}, {Nie}, {Ou}, {Qu}, {Sai}, {Shang}, {Shen}, {Sun}, {Tan}, {Tao}, {Tuo}, {Wang}, {Wang}, {Wang}, {Wang}, {Wang}, {Wang}, {Wang}, {Wen}, {Wu}, {Wu}, {Wu}, {Xiao}, {Xiong}, {Yan}, {Yang}, {Yang}, {Yang}, {Yi}, {Yuan}, {Zhang}, {Zhang}, {Zhang}, {Zhang}, {Zhang}, {Zhang}, {Zhang}, {Zhang}, {Zhang}, {Zhang}, {Zhang}, {Zhang}, {Zhang}, {Zhang}, {Zhang}, {Zhang}, {Zhang}, {Zhang}, {Zhang}, {Zhang}, {Zhao}, {Zhao}, {Zheng}, {Zhou}, {Zhu}, {Zhu}, {Zhuang}, \& {Insight-HXMT Team}}]{2020Zhang}
{Zhang}, S.-N., {Li}, T., {Lu}, F., {et~al.} 2020, Science China Physics, Mechanics, and Astronomy, 63, 249502, \dodoi{10.1007/s11433-019-1432-6}

\bibitem[{{Zimmerman} {et~al.}(2005){Zimmerman}, {Narayan}, {McClintock}, \& {Miller}}]{2005ApJ...618..832Z}
{Zimmerman}, E.~R., {Narayan}, R., {McClintock}, J.~E., \& {Miller}, J.~M. 2005, \apj, 618, 832, \dodoi{10.1086/426071}

\end{thebibliography}
\bibliographystyle{aasjournal}



\end{document}